\documentclass{Manuscript_Template_ICCA9_LaTex}
\usepackage{graphicx}
\usepackage{amsmath}
\usepackage{amssymb}
\usepackage{amsthm}
\usepackage{url}
\newcommand{\bma}{\left[\begin{matrix}}
\newcommand{\ema}{\end{matrix}\right]}
\newcommand{\be}{\begin{equation}}
\newcommand{\ee}{\end{equation}}
\newcommand{\BC}{\mathbb{C}}
\newcommand{\BR}{\mathbb{R}}

\newcommand{\bx}{\mathbf{x}}

\newcommand{\bn}{\mathbf{n}}
\newcommand{\diag}{\mathrm{diag}}
\newcommand{\im}{\mbox{Im}}
\newcommand{\Ker}{\mbox{Ker}}
\newcommand{\Tr}{\mbox{Tr}}
\newtheorem{thm}{Theorem}[subsection]

\newtheorem{lem}[thm]{Lemma}

\newtheorem{prop}[thm]{Proposition}

\title{CONFORMALLY COMPACTIFIED MINKOWSKI SPACE:
MYTHS AND FACTS}
\author{Arkadiusz Jadczyk}
\address{Center CAIROS, Institut de Math\'{e}matiques de Toulouse\\
Universit\'{e} Paul Sabatier, 118 Rout\'{e} de Narbonne 31062, Toulouse Cedex 9, France\\
\selectfont \normalfont
E-mail: arkadiusz.jadczyk@cict.fr}
\abstract{Maxwell's equations are invariant not only under the Lorentz group but also under the conformal group. Irving E. Segal has shown that while the Galilei group is a limiting case of the Poincar\'{e} group, and the Poincar\'{e} group comes from a contraction of the conformal group, the conformal group ends the road,  it is {\em rigid\,}. There are thus compelling mathematical and physical reasons for promoting the conformal group to the role of the fundamental symmetry of space--time, more important than the Poincar\'{e} group that formed the group-theoretical basis of special and general theories of relativity.
While the action of the conformal group on Minkowski space is singular, it naturally extends to a nonsingular action on the compactified Minkowski space, often referred to in the literature as ``Minkowski space plus light-cone at infinity''.  Unfortunately in some textbooks the true structure of the compactified Minkowski space is sometimes misrepresented, including false proofs and statements that are simply wrong.

In this paper we present in, a simple way, two different constructions of the compactified Minkowski space, both stemming from the original idea of Roger Penrose, but putting stress on  the mathematically subtle points and relating the constructions to the Clifford algebra tools. In particular the little-known antilinear Hodge star operator is introduced in order to connect real and complex structures of the algebra. A possible relation to Waldyr Rodrigues' idea of gravity as a plastic deformation of Minkowski's vacuum is also  indicated.}
\begin{document}
\section{Preliminaries}
\subsection{Notation \label{sec:not}}Let $E$ be a real 6--dimensional vectors space endowed with a bilinear form $(x,y)$ of signature $(4,2).$ Let $Q$ be the diagonal $6\times 6$ matrix \be Q=\mbox{diag }(1,1,1,-1,1,-1).\label{eq:q}\ee  We will call a basis $e_i$ in $E$ orthonormal if $(e_i,e_j)=Q_{i,j}.$\footnote{According to our conventions, in $E_{4,2},$ the first four coordinates $x^1,x^2,x^3$ and $x^4$ will correspond to Minkowski space coordinates $x,y,z$ and $t,$ while the coordinates $x^5,x^6$ will correspond to the added hyperbolic plane $E_{1,1}.$}  Any two orthonormal bases of $E$ are related by a transformation from the group $O(4,2):$
\be O(4,2)=\{R\in \mbox{Mat}(6,\BR):\,R\;Q\;{}^t\! R=Q\}.\ee When a preferred orthonormal basis is selected in $E,$ then $E$ is denoted by $E_{2,2}.$ For $x\in E_{4,2}$ we write \be Q(x)={}^t\!xQx=(x^1)^2+(x^2)^2+(x^3)^2-(x^4)^2+(x^5)^2-(x^6)^2.\ee
Let $G$ be the diagonal $4\times 4$ matrix $G=\diag (+1,+1,-1,-1).$ Let $V$ be a four--dimensional complex vector space endowed with a pseudo--Hermitian form $(\cdot|\cdot )$ of signature $(2.2).$ A basis $e_i$ of $H_{2,2}$ is said to be orthonormal if $(e_i|e_j)=G_{ij}.$
Any two orthonormal bases in $H_{2,2}$ are related by a transformation from the group $U(2,2):$
 \be U(2,2)=\{U\in \mbox{Mat}(4,\BC):UGU^*=G\}.\ee
 When a preferred orthonormal basis is selected in $V,$ then $V$ is denoted by $H_{2,2}.$ \\
\subsection{The algebras $Cl_{4,2}\approx \mbox{Mat}(4,\BR),$ $Cl^+_{4,2}\approx\mbox{Mat}(4,\BC),$ and the groups $SO_+(4,2)$  and $U(2,2)$}
Define the following six $4\times 4$ antisymmetric matrices $\Sigma_\alpha=(\Sigma_\alpha^{AB})$:
$$ \Sigma_1=\left(\begin{smallmatrix}
 0 & 0 & -i & 0 \\
 0 & 0 & 0 & i \\
 i & 0 & 0 & 0 \\
 0 & -i & 0 & 0
\end{smallmatrix}
\right)\, \Sigma_2=\left(
\begin{smallmatrix}
 0 & 0 & -1 & 0 \\
 0 & 0 & 0 & -1 \\
 1 & 0 & 0 & 0 \\
 0 & 1 & 0 & 0
\end{smallmatrix}
\right)\,
\Sigma_3=\left(
\begin{smallmatrix}
 0 & 0 & 0 & i \\
 0 & 0 & i & 0 \\
 0 & -i & 0 & 0 \\
 -i & 0 & 0 & 0
\end{smallmatrix}
\right)$$
$$\Sigma_4=\left(
\begin{smallmatrix}
 0 & i & 0 & 0 \\
 -i & 0 & 0 & 0 \\
 0 & 0 & 0 & -i \\
 0 & 0 & i & 0
\end{smallmatrix}
\right)\quad \Sigma_5=\left(
\begin{smallmatrix}
 0 & 0 & 0 & 1 \\
 0 & 0 & -1 & 0 \\
 0 & 1 & 0 & 0 \\
 -1 & 0 & 0 & 0
\end{smallmatrix}
\right)\quad
\Sigma_6=\left(
\begin{smallmatrix}
 0 & 1 & 0 & 0 \\
 -1 & 0 & 0 & 0 \\
 0 & 0 & 0 & 1 \\
 0 & 0 & -1 & 0
\end{smallmatrix}
\right)$$
The following identities hold:
\be \overline{\Sigma_\alpha^{ij}}=\frac{1}{2}\epsilon^{ijkl}G_{km}G_{ln}\Sigma_\alpha^{mn},\label{eq:sdsigma}\ee
where $\alpha=1,...,6;$ $i,j,k,l,m,n=1,...,4.$\\
\noindent
Recall that for an antilinear operator $A$ acting on a pseudo-Hermitian space $V$ the adjoint $A^*$ is defined by the formula:
$(Av|w)=(A^*w,v).$ The following proposition holds:
\begin{prop}
Define the following six complex matrices \be{\Gamma_\alpha^i}_k={\Sigma_\alpha}^{ij}G_{jk},\, (\alpha=1,...,6;\,i,j,k=1,...,4)\label{eq:cli}.\ee Explicitly:
$$ \Gamma_1=\left(\begin{smallmatrix}
 0 & 0 & i & 0 \\
 0 & 0 & 0 & -i \\
 i & 0 & 0 & 0 \\
 0 & -i & 0 & 0
\end{smallmatrix}
\right)\, \Gamma_2=\left(
\begin{smallmatrix}
 0 & 0 & 1 & 0 \\
 0 & 0 & 0 & 1 \\
 1 & 0 & 0 & 0 \\
 0 & 1 & 0 & 0
\end{smallmatrix}
\right)\,
\Gamma_3=\left(
\begin{smallmatrix}
 0 & 0 & 0 & -i \\
 0 & 0 & -i & 0 \\
 0 & -i & 0 & 0 \\
 -i & 0 & 0 & 0
\end{smallmatrix}
\right)$$
$$\Gamma_4=\left(
\begin{smallmatrix}
 0 & i & 0 & 0 \\
 -i & 0 & 0 & 0 \\
 0 & 0 & 0 & i \\
 0 & 0 & -i & 0
\end{smallmatrix}
\right)\quad \Gamma_5=\left(
\begin{smallmatrix}
 0 & 0 & 0 & -1 \\
 0 & 0 & 1 & 0 \\
 0 & 1 & 0 & 0 \\
 -1 & 0 & 0 & 0
\end{smallmatrix}
\right)\quad
\Gamma_6=\left(
\begin{smallmatrix}
 0 & 1 & 0 & 0 \\
 -1 & 0 & 0 & 0 \\
 0 & 0 & 0 & -1 \\
 0 & 0 & 1 & 0
\end{smallmatrix}
\right).$$
Let $\Gamma_\alpha$ be the antilinear operators on $H_{2,2}$ defined by the formula:
$$({\Gamma_\alpha} f)^i ={ \Gamma_\alpha^i}_j\,\overline{f^j},\quad f=(f^i)\in H_{2,2}.$$ Then the antilinear operators $\Gamma_\alpha$ are anti--self--adjoint: $\Gamma_\alpha=-\Gamma_\alpha^*,$ and satisfy the following anti--commutation relations of the Clifford algebra of $E^{4,2}:$
$$ \Gamma_\alpha\circ\Gamma_\beta +\Gamma_\beta\circ\Gamma_\alpha = 2\,Q_{\alpha_\beta}.$$
The space $H_{2,2}$ considered as an $8$--dimensional {\bf real\,} vector space carries this way an irreducible representation of the Clifford algebra $Cl_{4,2}.$ The Hermitian conjugation in $H_{2,2}$ coincides with the conjugation of $Cl_{4,2}.$ The space $H_{2,2}$ considered as a $4$-dimensional {\bf complex\,} vector space carries a faithful irreducible representation of the even Clifford algebra $Cl^{+}_{4,2}.$\\
For each $x=(x^1,...,x^6)\in E_{4,2},$ let $X$ be the matrix
\be X=\sum_{\alpha=1}^6 x^\alpha \Gamma_\alpha,\label{eq:X}\ee
then
\be
X=\begin{pmatrix}
 0 & i x_4+x_6 & i x_1+x_2 & -i x_3-x_5 \\
 -i x_4-x_6 & 0 & -i x_3+x_5 & -i x_1+x_2 \\
 i x_1+x_2 & -i x_3+x_5 & 0 & i x_4-x_6 \\
 -i x_3-x_5 & -i x_1+x_2 & -i x_4+x_6 & 0
\end{pmatrix}\ee
We have \be
\bar{{X}^i}_j=\frac{1}{2}\epsilon^{imnk}\,G_{mj}G_{nl}\;{X^l}_k,\ee
and
\be \det (X)=\det (\bar{X})=Q(x)^2,\ee
where $\bar{X}$ is the complex conjugate matrix.
\label{prop:cl42}\end{prop}
\footnote{
There are several errors in Ref. \cite{aj1}: in the listing of the matrices $\Sigma_\alpha,$ the matrix $\Sigma_2$ is listed twice; in Lemma 5, Eq. (31) instead of $Q_{jk}$ should be $G_{jk};$ in a formula below, $\Gamma^\alpha$ should be $\Gamma_\alpha;$ it is stated that the Hermitian conjugation coincides with the main antiautomorphism, which is incorrect, it should be `conjugation'. Also the Lorentz Lie algebra block matrix at the end of section 8.3 should be in the upper left corner.}

We recall \cite[p. 387, Definition IX.4.C]{deheuvels} that the group $Spin(E)$ consist of products of even numbers of vectors $x\in E$ with $Q(x)=1,$ and even numbers of vectors $y\in E$ with $Q(y)=-1.$ The action of $Spin(E)$ on $E\subset Cl(4,2)$ is given by $x\mapsto \pi(g)xg^{-1},$ which is the same as $gxg^{-1}$ when $g\in Spin(E)\subset Cl^+_{4,2}.$\footnote{We denote by $\pi$ (resp. by $\tau$) the main automorphism (resp. antiautomorphism) of the Clifford algebra.} If $x,x'$ are two normalized  in $E,$ then their product $xx'$ operates on $H_{2,2}$ via a {\bf complex linear} transformation implemented by  the matrix $X\bar{X'}$ of determinant one. It follows that, with our identification, the group $Spin(E)$ coincides with the group $SU(2,2).$ In fact, for $x\in E_{4,2}$ and $U\in SU(2,2,$ we have
$ UXU^*=X',$ where
\be X'^\alpha = {L(U)^\alpha}_\beta\;X^\beta,\ee
where $U\mapsto L(U)$ is a homomorphism from $SU(2,2))\approx Spin(4,2)$ onto $SO_+(E_{4,2})$ with kernel $\{1,-1,i,-i\}.$
\section{The exterior algebra $\Lambda^2H_{2,2}$}
Let $\Lambda^2H_{2,2}$ be the (complex) exterior algebra of $H_{2,2}.$ It carries a natural pseudo-Hermitian form:
\be (x|y) = \frac{1}{p!}\;G_{i_1j_1}..G_{i_p j_p}x^{i_1...i_p}\;\overline{y^{j_1...j_p}}.\label{eq:xy}\ee If $e_i$ is the orthonormal basis of $H_{2,2},$ we define the following six bivectors $E_\alpha\in\Lambda^2H_{2,2}:$
\be E_\alpha=\frac{1}{2\sqrt{2}}\,\Sigma_\alpha^{ij}\,e_i\wedge e_j\ee They are normalized:
\be (E_\alpha|E_\beta)=Q_{\alpha\beta}.\ee
In \cite{sto} an antilinear Hodge $\star$ operator has been defined in the case of the Euclidean signature. It can be readily extended to the pseudo--Euclidean case. Thus we define $\star:\;\Lambda^k H_{2,2}\rightarrow \Lambda^{4-k}H_{2,2}$ by the formula:
\be x\wedge \star y = (x|y)e,\quad x,y\in\bigwedge^k\, V ,\label{eq:sto}\ee
where $e=e_1\wedge...\wedge e_4,$ $x\in \Lambda^kH_{2,2},$ $y\in\Lambda^{4-k}H_{2,2}.$ Notice that we have:
\be (x|\star y)=(-1)^{k(4-k)}(y|\star x),\label{eq:star}\ee
We easily find that
$\star\star\, x=(-1)^{k(n-k)}x,\; x\in \Lambda^kH_{2,2}.$ It follows that on $\Lambda^2H_{2,2}$ we have $\star^2=1.$  Thus $\Lambda^2 H_{2,2},$ which is a vector space of complex dimension $6$ splits into a direct sum of two \textbf{real} $6$--dimensional subspaces
\be \Lambda^2H_{2,2}=\Lambda^2_+H_{2,2}\oplus\Lambda^2_-H_{2,2},\ee where \be \Lambda_\pm ^2H_{2,2}=\{x\in\Lambda^2H_{2,2}:\star x=\pm x\}.\ee The multiplication by the imaginary unit $i$ gives a bijective correspondence between these subspaces. The bivectors $E_\alpha$ are self--dual: $\star E_\alpha=E_\alpha,$ and form a basis in  $\Lambda_+^2H_{2,2}.$

\textbf{Thus $E_{4,2}$ can be interpreted in two ways: either as a real linear subspace of anti--linear transformations of $H_{2,2}$ determined by the matrices of the form (\ref{eq:X}), or as the subspace $\Lambda^2_+H_{2,2}$ of self--dual bivectors in $\Lambda^2H_{2,2},$ as suggested originally by Kopczy\'{n}ski and Woronowicz \cite{wk}, though these authors did not recognize the relation of their reality condition to the anti--linear Hodge operator (\ref{eq:star}).}
\subsection{Compactified Minkowski space}
Early in the XX-th century Bateman and Cunningham \cite{bateman1,cunningham1,bateman2} established invariance of the wave equation and of Maxwell's equations under conformal transformations. The central role in these transformations is being played by the conformal inversion, formally defined by \be R:\, (\bx,t)\mapsto \frac{(\bx,t)}{\bx^2-c^2t^2}.\label{eq:ci}\ee It is singular on the light cone $x^2=\bx^2-c^2t^2=0.$
More general, special conformal transformations, of the form $RT(a)R,$ where $T(a)$ is the translation by a vector $a$ in $\BR^4,$ are also singular in the Minkowski space--time $M.$ In order to avoid singularities the conformally compactified Minkowski space $M^c$ - a homogeneous space for the conformal group $SO_+(4,2)$ is introduced. There are three different, though related to one another, ways of describing $M^c:$ a) The group manifold of the unitary group $U(2),$ b) The projective quadric defined by the equation $Q(x)=0$ in $E_{4,2},$ and c) The space of maximal isotropic subspaces of $H_{2,2}.$ Let us discuss briefly the relation between b) and c). The relation to a) has been discussed in \cite{aj1}.\\
\noindent Let $\varphi$ be the mapping $\varphi:\, E_{4,2}\rightarrow \Lambda_+H_{2,2}$ defined by
$\varphi(x)=x^\alpha E_\alpha.$
The following lemma holds \cite{aj1}.
\begin{lem}
The mapping $\varphi$ defines a bijective correspondence between generators of the null cone $Q(x)=0$ in $E_{4,2}$ and maximal isotropic subspaces in $H_{2,2}.$
\end{lem}
In particular, $Q(x)=0,$ if and only if $\varphi(x)$ is a bivector of the form $v\wedge w,$ where $v$ and $w$ are two linearly independent, mutually orthogonal, isotropic vectors in $H_{2,2}.$ There is another method of viewing this correspondence, studied by Ren\'{e} Deheuvels in \cite[Th\'{e}or\`{e}me VI.6.D, p. 283]{deheuvels1}, and generalized by Pierre Angl\`{e}s \cite{angles}. We start with a brief recapitulation, with a slight change of notation, of the construction given in \cite{angles}. The construction is quite general. but we restrict ourselves to the case we are interested in. Let $E=E_{4,2},$ $V=H_{2,2}.$  It is convenient to identify $x\in E$ with the antilinear operator $\varphi(x)$ acting on the spinor space $V.$ Then the following  theorem holds:
\begin{thm}{\normalfont{\cite[1.5.5.1.1, p. 45]{angles}}}
The mapping
\begin{quotation} $\{$ isotropic line $\BR x$ of $E_{4,2}\}$ $\mapsto$ maximal totally isotropic subspace $S(x)$ of $V$ is injective and realizes a natural embedding of $\tilde{Q}(E),$ the projective quadric associated with $E,$ into the Grassmannian $G(V,\frac{1}{2}\dim V)$ of subspaces of $V$ of dimension $\frac{1}{2}\dim\,V.$
\end{quotation}
\end{thm}

The proof goes as follows: given a null vector $x\in E\subset L_\BR(V)\approx Cl_{4,2},$ there exists another null vector $y\in E$ such that $(x,y)=2.$ Denoting by $x,y$ antilinear operators on $V$ representing $x,y,$  in $L_\BC(V),$ $(xy)^2=xy,$ $(yx)^2=yx,$ and $xy+yx=I.$ Therefore $xy$ and $yx$ are two complementary idempotents in $L_\BC(V).$ Evidently we have $\im(xy)\subset \im(x).$ Since $x^2=0$ we have that $\im(x)\subset \Ker(x).$ On the other hand, if $v\in V$ is in $\Ker(x)$ then $v=(xy+yx)v=xyv,$ therefore $\Ker(x)\subset \im(xy)\subset \im(x).$ It follows that $\im(x)=\im(xy)=\Ker(x).$ Now, owing to the fundamental property of the trace, we have $\Tr_\BR(xy)=\Tr_\BR(yx).$ But since $xy$ and $yx$ are complex linear, it follows that $\Tr_\BC (xy)=\frac{1}{2}\Tr_\BR(xy)=\frac{1}{2}\Tr_\BR(yx)=\Tr_\BC (yx).$ Since  $xy$ and $yx$ are complementary idempotents, it follows that $\Tr_\BC (xy)=\dim_\BC (\im(xy))=\dim_\BC (\im(yx)).$ Then, from $\im(xy)\oplus \im(yx)=V,$ $\im(xy)\cap \im(yx)=\{0\},$ we deduce that $\dim(\im(xy))=\dim(\im(yx))=\frac{1}{2}\dim (V).$ In our particular case $\frac{1}{2}\dim (V)=2.$

In order to complete the demonstration we notice that $(xy)^\tau=(xy)^*=yx$ therefore for all $s,t\in V$ we have
$(xy\,s|xy)=(s,(xy)^*yy|t)=(s|yxxy\,t)=0.$ It follows that $\im(xy)$ is a totally isotropic subspace, thus maximal totally isotropic. On the other hand from the equality $\im(xy)=\im(x)$ we find that this subspace is independent of the choice of $y$ having the required properties.

In our case the correspondence $x\mapsto S(x)$ is not only an embedding, but also a \textbf{bijection}. This follows form the construction above, the covering homomorphism $U(2,2)\rightarrow O(4,2)$ and the fact that the pseudo--orthogonal group $O(4,2)$ (resp. unitary group $U(2,2)$) acts transitively on totally isotropic subspaces of $E$ (resp. of $V$) of the same dimension - cf. \cite[Corollaire 2, p. 74]{b9}

While isotropic lines $\BR x$ in $E$ (or maximal isotropic subspaces $S(x)$ in $V$) correspond to the points of the compactified Minkowski space $M^c,$ there is an interesting duality: maximal isotropic subspaces (isotropic planes) in $E$ are in one--to--one correspondence with isotropic (complex) lines in $V,$ and they correspond to null geodesics in $M^c.$\footnote{It is important to notice that naturally $M^c$ carries only a conformal structure rather than a Riemannian metric. But {\textbf null} geodesics are the same for each Riemannian metric in the conformal class.}
\subsection{From maximal isotropic subspaces in $E$ to isotropic lines in $V.$}
We will use the method and the results of the previous subsection. Let $N$ be a maximal isotropic subspace of $E.$ Since, in our case, $E=E_{4,2},$ it follows that $N$ is two--dimensional. It is then known \cite[p. 77-78]{chevalley} that there exists another maximal isotropic subspace $P$ such that $N\cap P=\{0\},$ vectors $x_1,x_2$ spanning $N$ and vectors $y_1,y_2$ spanning $P,$ with the property $(x_i,y_j)=2\delta_{ij},\quad i=1,2.$ We define $P_i=x_iy_i,$ $Q_i=y_ix_i,$ and it easily follows that $P_i^2=P_i,$ $Q_i^2=Q_i,$ $P_i+Q_i=I,$ and, moreover, $P_i$ and $Q_j$ commute for $i\neq j.$ It follows $R_1=P_1P_2,R_2=P_1Q_2,R_3=Q_1P_2,R_4=Q_1Q_2$ are four idempotents with $R_iR_j=0,\quad i=1,2,3,4,$ and $R_1+R_2+R_3+R_4=I.$ It is easy to see that $\Tr(R_1)=...=\Tr(R_4).$ For instance \begin{eqnarray*}\Tr(R_1)&=&\frac{1}{2}\Tr_\BR (x_1y_1x_2y_2)=\frac{1}{2}\Tr_\BR (y_2x_1y_1x_2)=\frac{1}{2}\Tr_\BR(x_1y_1y_2x_2)\\&=&\Tr_\BC(P_1Q_2)=\Tr(R_2),\end{eqnarray*} where we have use the fact $y_2$ anticommutes with $x_1$ and $y_1,$ therefore commutes with $x_1y_1.$ Therefore we have that $\dim \im(R_i)=1.$ Since $\im(R_1)\subset \im(x_1),$ the subspace $\im (R_1)$ is an isotropic line.\\
Let us now show that $\im (R_1)$ depends only on the subspace $N$ and not on the choice of the auxiliary subspace $P$ or a particular choice of our vectors $x_i$ and $y_i.$ For this end we first notice that $\im (R_1)=\im (x_1 x_2).$ Indeed, we evidently have $\im (x_1y_1x_2y_2)\subset\im (x_1x_2).$ On the other hand assume that $s\in \im(x_1x_2),$ then, for some $t\in V,$ we have
$$s=x_1x_2\,t =x_1x_2(x_1y_1x_2y_2+x_1y_1y_2x_2+y_1x_1x_2y_2+y_1x_1y_2x_2).$$
Multiplying, using commutation properties and nilpotency of $x_i,$ we find that only the last term survives and it can be written as
$x_1x_2x_2y_1x_1y_2x_2\,t= x_1y_1x_2y_2(x_1x_2\,t).$ Therefore $s\in \im (R_1),$ and so we have shown that $S(x_1,x_2)\stackrel{\mbox{df}}{=}\im(R_1)=\im(x_1x_2)$ does not depend on the choice of $y_i$ with the properties as above. Finally, if $x'_i=\sum_{j=1}^2a_{ij}x_j$ is a nonsingular transformation of the basis $x_i$ of $N,$ then a simple calculation gives that
$x_1'x_2'=\det(a)\,x_1x_2,$ and therefore $\im(x_1x_2)=\im(x_1'x_2').$

There is an alternative way of looking at this correspondence: let $v$ be a non--zero null vector in $V,$ and let $N(v)$ be defined as:
\be N(v)=\{x\in E:\, xv=0\}.\ee
Then $N(v)$ is a maximal isotropic subspace of $E$ and $\BC v$ corresponds to $N(v)$ according to the construction above. Indeed, it follows immediately from the construction that $S(x_1,x_2)\subset N(v).$ To show that, in fact, equality holds, it is enough to show that $N(v)$ is an isotropic subspace of $E.$ If $x^2\neq 0,$ and $x\in N(v),$ then for all $w\in V$ we have $0=(xv|xw)=-(w|x^2v),$ and therefore we must have $x^2=0.$

\subsection{$E_{4,2}$ as the arena for Lie spheres of $\BR^3$}
The space $E_{4,2}$ carrying a natural linear representation of the conformal group $O(4,2)$ has been introduced by Sophus Lie in 1872 \cite{lie1}, and developed further by W. Blaschke in 1929 \cite{blaschke3}. We will follow the modern presentation of Ref. \cite{cecil1}, though we will change the coordinate labels so as to adapt them to our notation introduced in section \ref{sec:not}. It is instructive to see how the Lorentz, Poincar\'{e} and the conformal  group naturally enter the scene without any philosophical load of Einstein's relativity.\\
\noindent We can assume absolute time. We can also assume absolute space, its points represented by coordinates of $\BR^3,$ and compactified by adding one point: $\infty,$ and absolute time, represented by points of $\BR.$ An oriented sphere with center at $ \bx$ and radius $r$ can be also interpreted as the coordinate of an 'event' in space--time, taking place at $\bx$ at time $r/c.$\footnote{We will assume the system of units in which $c,$ the speed of light, is numerically equal to $1.$} We allow for $r$ to be negative, with the interpretation that negative radius corresponds to the negative orientation of the sphere. The radius $r$ can be interpreted as the radius of a spherical wave at time $t$, if the wave, propagating through space with the speed light, was emitted at $\bx$ at time $t=0.$ The image being that when the spherical wave reduces to a point, it turns itself inside--out, thus reversing its orientation. As a limit case there will also be spheres of infinite radius - represented by oriented 2--planes - these will correspond to events that took place in infinitely distant past or future. Points (spheres of zero radius), spheres, and planes (spheres of infinite radius) are bijectively represented by generator lines of the null cone in $E_{4,2}$ as follows (cf. \cite[p. 16]{cecil1}):\vskip12pt
\begin{center}
\begin{tabular}{cc}
\textbf{Euclidean}&\textbf{Lie}\\
points: $\bx\in\BR^3$&$[(\bx,0,\frac{1+\bx^2}{2},-\frac{1-\bx^2}{2},0,0)]$\\
$\infty$& $[(\mathbf{0},0,1,1)]$\\
spheres: center $\bx,$ signed radius $t$&$[(\bx,t,\frac{1+\bx^2-t^2}{2},-\frac{1-\bx^2+t^2}{2}$)]\\
planes: $\bx\cdot \bn=h,$ unit normal $\bn$&$(\bn,1,h,h)]$\\
\end{tabular}
\end{center}
\begin{center}Table 1: Correspondence between Lie spheres and points of the compactified Minkowski space. $[x]$ denotes the equivalence class modulo $\BR^*.$
\end{center}
\vskip12pt
\section{Myths and facts}
One of the `myths' we have already encountered above. Minkowski space and its compactification arise naturally through the studies of geometry of spheres in $\BR^3$ - it not necessary to invoke Einstein's relativity principle, or restrict the range of available velocities, as it is usually being done. Second myth, more serious one, can be summarized in one sentence: `conformal infinity' is the result of the conformal inversion of the light cone at the origin of $M.$' Such a statement can be found, for instance, in \cite[p. 127]{akivis}. Conformal inversion is implemented by $O(4,2)$ transformation $(\mathbf{x},t,v,w)\mapsto (\mathbf{x},t,-v,w).$ Conformal infinity of $M$ is represented by generator lines of the quadric $Q(x)=0$ of the form $(\mathbf{x},t,v,v).$ According to Table 1, the light cone $\mathbf{x}^2=t^2$ is represented in $E$  by generator lines of the form $[(\mathbf{x},t,1/2,-1/2)].$ Applying conformal inversion we get $[(\mathbf{x},t,-1/2,-1/2)].$ \textbf{Clearly the whole 2-sphere $S^2$ of $[(\mathbf{x},1,0,0)],$ $\mathbf{x}^2=1$ is missing. This two--sphere is located at conformal infinity and is pointwise invariant under the conformal inversion.} The third myth is closely related, and it is usually summarized by giving to conformal infinity the name `light cone at infinity' \cite{penrose1} or, sometimes, `double light cone at infinity'. For instance Huggett and Tod write \cite[p. 36]{tod}: \textit{`Thus $M^c$ consists of $M$ with an extra null cone added at infinity.'\,} Not only they write so in words, but also miss this $S^2$ in their formal analysis.  A pictorial representation of the conformal infinity (suppressing just one dimension) is that of one of the degenerate cases of Dupin cyclides\footnote{The exact connection of these two concepts is not known to the present author at the time of this writing.} - so called \textit{needle (horn) cyclide\,} \cite[Fig. 6, p. 80]{schrott}, \cite[Fig. 5.7, p. 156]{cecil1}, or, in French, \textit{croissant simple\,} \cite{ferreol}:
\begin{figure}[!ht]
\begin{center}
      \includegraphics[width=5cm, keepaspectratio=true]{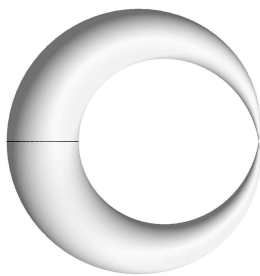}
\end{center}
  \caption{Pictorial representation of the conformal infinity with one dimension skipped. \textit{`Double light cone at infinity'\,}, with endpoints identified. While topologically correct such a name is misleading as it suggests non differentiability at the base, where the two half-cones meet - cf. \cite{aj1}. The ill-fated sphere $S^2$ is marked.}
\label{fig:fig3}\end{figure}
\section{Concluding comments}We discussed, briefly, several interesting properties of the compactified Minkowski space - an important homogeneous space for the conformal group, the group of symmetries of Maxwell equations discovered long ago. Conformal group appears to be important in several areas of physics, mathematics, also in computer graphics and pattern recognition. It is quite possible that its full potential has not yet been exhausted. One of the interesting properties of the conformal group is that it is the first element in a sequence of group contractions \cite{wigner}: conformal group, Poincar\'{e} group, Galilei group. The Lie algebra of the conformal group,  as demonstrated by I. Segal in the final part of his 1951 paper \cite{segal1}\footnote{Cf. also \cite{segal1a}}, does not result as a limit of some other Lie algebra. Segal has constructed the foundations of his cosmological model \cite{segal2} starting from the compactified Minkowski space and its universal covering space, a model with some difficulties, but also with some promises if connected with elementary particle physics (cf. \cite{segal3}, (see also R. I, Ingraham, via a somewhat different, but related path \cite{ingraham1}, and a recent paper extending the idea of Segal's chronogeometry in new directions, by A. V. Levichev \cite{levichev}). Segal himself did not really touch the problem of placing gravity within his framework. But recent papers (cf. \cite{levichev} and references therein) seem to point at the possibility of realizing gravity along the lines suggested also by Waldyr Rodrigues in his monograph written with V. V. Fern\'{a}ndez \cite{wr1}: gravitation is  somehow related to  `quantum fluctuations of the vacuum'. The geometrical arena for such a description can be either Minkowski space or, what seems to be more attractive to the present author, one of the homogeneous spaces of the conformal group (or its covering).
\section{Acknowledgements}
The author thanks Pierre Angl\`{e}s for reading the paper, his  kind encouragement and advice, and for constructive and critical comments.

\end{document}